\documentclass[
 article,
superscriptaddress,
twocolumn,
 amsmath,amssymb,
 aps,
]{revtex4-2}

\usepackage{graphicx}
\usepackage{dcolumn}
\usepackage{bm}
\usepackage{tikz}
\usepackage[compat=1.1.0]{tikz-feynman}
\usepackage{mathtools}
\begin{document}



\preprint{APS/123-QED}

\title{Intrinsic surface superconducting instability in Type-I Weyl Semimetals}

\author{Aymen Nomani}
\affiliation{Department of Physics, University of Houston, Houston, TX 77204, USA}
\author{Pavan Hosur}
\affiliation{Department of Physics, University of Houston, Houston, TX 77204, USA}
\affiliation{Texas Center for Superconductivity, University of Houston, Houston, TX 77204, USA}

\date{\today}

\begin{abstract}
 Recent experiments on non-magnetic Weyl semimetals have seen separate bulk and surface superconductivity in Weyl semimetals, which raises the question of whether the surface Fermi arcs can support intrinsic superconductivity while the bulk stays in the normal state. A theoretical answer to this question is hindered by the absence of a well-defined surface Hamiltonian since the Fermi arcs merge with the bulk states at their endpoints. Using an alternate, Green's functions-based approach on a phenomenological model that can yield arbitrary Fermi arcs, we show -- within mean-field theory -- that the surface can support a standard Cooper instability while the bulk remains disordered. Although the surface has lower dimensionality, a higher density of states compared to the bulk allows it to have a higher mean-field superconducting transition temperature.

\end{abstract}

\maketitle

\section{Introduction}

Weyl semimetals are three-dimensional (3D) topological materials defined by the presence of non-degenerate bands that intersect at discrete points in bulk momentum space \cite{VafekDiracReview,Burkov2018,Burkov:2016aa,YanFelserReview,Armitage2018,Shen2017,Belopolski:2016wu,Guo2018,Chang2016,Gyenis_2016,Huang:2015vn,Inoue1184,Lv:2015aa,Sun2015a,Xu2015,Xu2016,Xu613,Yang:2015aa,Zheng2016}. These points are known as Weyl nodes because the low energy dispersion around them resembles that of a Weyl fermion. Weyl nodes have a well-defined chirality or handedness and occur in even numbers in a Weyl semimetal, with half of each chirality. They also carry topological protection in the sense they cannot be gapped out perturbatively while translational symmetry of the material persists; when it does not, they can only be annihilated in pairs of opposite chirality \cite{Hosur2013a,Wang_2018,Hu:2019aa,ZyuninBurkovWeylTheta,ChenAxionResponse,VazifehEMResponse,Burkov_2015,Hosur2012,Juan:2017aa,Wang2017,Halterman2018,Halterman2019,Nagaosa:2020aa,NielsenABJ,IsachenkovCME,SadofyevChiralHydroNotes,Loganayagam2012,GoswamiFieldTheory,Wang2013,BasarTriangleAnomaly,LandsteinerAnomaly}.

Recent years have seen significant experimental developments in the interplay of superconductivity with Weyl semimetallicity  \cite{WANG2017425, Aggarwal:2017aa, Wang2020, Luo2019, Li:2017ab, Delft2020, Kumar2016, Baenitz2019, Kang:2015aa, Pan:2015aa, Li2018, Qi:2016aa, Xing204, Cai2019, Mu2021, Mandal2021, Shipunov2020, Schimmel2023}. In type-I Weyl semimetal t-PtBi\textsubscript{2}, transport measurements on bulk single crystals showed superconductivity with a $T_C$ of $0.6K$ \cite{Shipunov2020} while scanning tunneling spectroscopy on the surface revealed a wide range of superconducting gaps, with the largest gaps corresponding to $T_C$ in the 100 K range \cite{Schimmel2023}. Powdered NbP was also found to exhibit superconductivity \cite{Kumar2016, Baenitz2019} with a small superconducting volume fraction, and Ref. \cite{Baenitz2019} speculated that the superconductivity could be occurring on the surface. This, along with a large difference in surface and bulk transition temperatures in t-PtBi\textsubscript{2}, raises the question, "Can the surface turn superconducting while the bulk remains in the normal state?"

The answer is hindered by another fundamental and exotic property of Weyl semimetals, namely, surface states known as the Fermi arc. These are open strings of zero energy states on the surface of a Weyl semimetal that connect the surface projections of Weyl nodes of opposite chirality. Unlike the Fermi surfaces of a conventional 2D metal, they do not form a closed contour; unlike the surface states of topological insulators, their penetration depth into the bulk depends strongly on the surface momentum and diverges at the end-points, causing their wavefunction to merge with the bulk Bloch waves at the Weyl nodes \cite{Benito-Matias2019,Chang2016,Gyenis_2016,Deng2017,Deng:2016aa,Guo2018,HaldaneFermiArc,Hosur2012a,Huang2016,Huang:2015vn,Iaia:2018aa,Inoue1184,Kwon:2020aa,Lau2017,Sakano2017,Sun2015a,Lv:2015aa,Xu2015,Xu2015a,Xu2015b,Xu2016,Xu613,XuLiu2018,Yuan2018QPI,Yuaneaaw9485,Zhang:2017ac,Moll:2016aa,Potter2014,Zhang2016}. The inseparability of the bulk and the surface makes it impossible to define a surface Hamiltonian, which hinders theoretical inquiries into the surface physics of Weyl semimetals. Nonetheless, the question raised above can be rephrased as "Does the Fermi arc metal support an intrinsic Cooper instability independently of the bulk?"

In this work, we explore the superconducting instability of the surface of time-reversal symmetric Weyl semimetals (TWSMs), since time-reversal symmetric Fermi surfaces generically have a superconducting instability, and discover an affirmative answer to the above question. Such an answer directly contrasts naïve expectations from Bardeen-Cooper-Schrieffer theory \cite{Bardeen1957}. According to the theory, higher dimensionality suppresses fluctuations and stabilizes mean-field superconductivity, suggesting that the bulk of a TWSM should be more susceptible to superconductivity than the surface. However, we find that the surface can turn superconducting before the bulk does. This is because the surface has a finite density of states due to the Fermi arcs, whereas the bulk density of states vanishes in the Weyl limit and remains parametrically small for a slightly doped Weyl node.

In Sec. \ref{sec:formalism}, we introduce the general Hamiltonian for the Weyl semimetal and show how Green's function formalism bypasses the problem of surface-bulk inseparability.  The interaction is introduced, which is an intra-layer Hubbard interaction with pair hopping. This interaction is then used to calculate the correlation function that induces a surface superconducting instability. The problem then reduces to calculating the second-order bubble diagram.  In Sec. \ref{layeredmodel}, we introduce a model of a TWSM with an arbitrary number and shape of Fermi arcs and the associated Green's function. Using this model, the contribution to the correlation function splits into two parts: the Fermi arcs and the projection of the bulk Fermi surface on the surface.  The contribution due to Fermi arcs is then calculated in Sec.  \ref{FermiArc}, and in Sec. \ref{sec:bulk projection}, we show the contribution due to the projection of the bulk states onto the surface.  In Sec. \ref{bulk instability}, we calculate the bulk instability.  Finally, in Sec. \ref{experiment}, we discuss the implications of our result in the context of the experiments performed on NbP and t-PtBi\textsubscript{2}. 

\section{General formalism}\label{sec:formalism}
In this section, we develop the formalism for studying the surface superconducting instability in a Type-I Weyl semimetal. While a surface Hamiltonian is ill-defined, a surface Green's function is meaningful and is the building block of our theory. Such an approach has previously been successful in evading this problem of bulk-surface inseparability and studying surface physics such as Friedel oscillations \cite{Hosur2012a}, conductivity \cite{Pal2022}, and Luttinger arcs \cite{Obakpolor2022}.  

We begin by considering a slab of a time-reversal symmetric Weyl semimetal described by the Bloch Hamiltonian $H_\textbf{k}$. We assume $2 D_z $ degrees of freedom in the $z^{th}$ layer -- the evenness mandated by time-reversal symmetry -- and decompose $H_\textbf{k}$ into blocks capturing the surface, the bulk and the surface bulk-coupling:
\begin{align}
H_\textbf{k} = \begin{pmatrix}
                            H_{\textbf{k}}^S & h_\textbf{k}\\
                             h_\textbf{k}^{\dagger} & H_\textbf{k}^B\\
\end{pmatrix}
\end{align}
Here, $H_\textbf{k}^S$ is the $2D_S\times2D_S$ in-plane Bloch Hamiltonian of the $z=0$ surface layer, $H_\textbf{k}^B$ is the Bloch Hamiltonian of all the other layers that we collectively refer to as ``bulk", while $h_\textbf{k}^{\dagger}$, $h_\textbf{k}$ capture the coupling between the bulk and the surface. The coupling terms can be strong, making it difficult to write an effective surface Hamiltonian, but an effective surface Green's function can be written. Specifically, writing Matsubara Green's function for the full slab in block form and evaluating the $2D_S$-dimensional block corresponding to the surface degrees of freedom yields an effective surface Green's function \cite{Obakpolor2022}

\begin{align}
g_{\textbf{k},i\omega_n} = (i\omega_n - H^S_\textbf{k} - h_\textbf{k}G^B_{\textbf{k}, i\omega_n} h_\textbf{k}^{\dagger})^{-1} 
\end{align}
where $G^B_{\textbf{k}, i\omega_n}=(i\omega_n-H^B_{\textbf{k}})^{-1}$. $g_{\textbf{k},i\omega_n}$ can alternately be obtained by integrating out the bulk fermions from a Euclidean path integral, see Appendix \ref{sgf}. Importantly, $g_{\textbf{k},i\omega_n}$ can be calculated analytically for certain local hopping models, as we demonstrate shortly.
 
 Next, in anticipation of deriving a large-$D_S$ mean-field theory, we introduce local, intra-layer attractive Hubbard and pair-hopping interactions that are invariant under $O(D_z)$ rotations within each layer. Explicitly,

\begin{align}
 H_{int} =  -  \sum_{\textbf{r},z} \sum_{n_z, n_z^\prime}  
\frac{U}{D_z} &c^\dagger_{\uparrow,\textbf{r} , z,  n_z} 
 c^\dagger_{\downarrow,\textbf{r}, z, n_z}\times \\
 &c_{\downarrow,\textbf{r}, z,  n_z^\prime }
 c_{\uparrow,\textbf{r}, z,  n_z^\prime}  \nonumber   
\end{align}
where $U>0$, $n$ indicates the orbital index, and $\textbf{r}$ is the 2D position vector. Fourier transforming in-plane,
\begin{align}
H_{int} =  -  \intop_{\textbf{k}',\textbf{k},\textbf{K}} \sum_{n_z, n_z^\prime} \sum_z  
\frac{U}{D_z} &c^\dagger_{\uparrow,\frac{\textbf{K}}{2}+\textbf{k}, z,  n_z} 
 c^\dagger_{\downarrow, \frac{\textbf{K}}{2}-\textbf{k}, z , n_z}\times \\
 &c_{\downarrow,\frac{\textbf{K}}{2}-\textbf{k}', z , n_z^\prime}
 c_{\uparrow,\frac{\textbf{K}}{2}+\textbf{k}', z , n_z^\prime}  \nonumber
\end{align}
where $\textbf{K}$, $\textbf{k}$ and $\textbf{k}'$ are 2D momenta and $\intop_\textbf{k}\equiv\int\frac{d^2k}{(2\pi)^2}$.

\begin{figure}[h]
\includegraphics[width=0.95 \columnwidth]{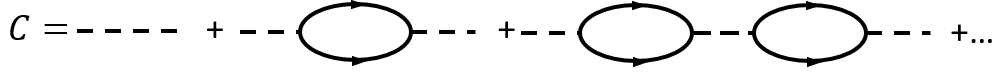}
\caption{\label{fig:fd} The Dyson series for $C_{\textbf{K},i\nu_n}$ in the large-$D_S$ limit. Superconductivity occurs when $C_{0,0}$ diverges.}
\label{fig:Dyson}
\end{figure}

\begin{figure}[h]
\includegraphics[width=0.95\columnwidth]{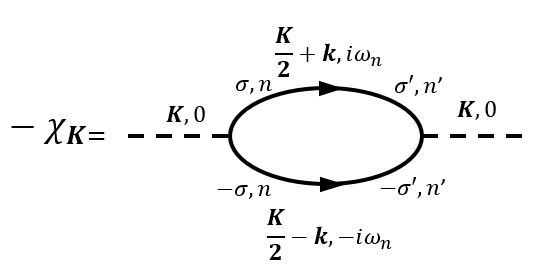}
\caption{\label{fig:fd} The second order bubble diagram $-\chi_{\textbf{K}}$ helps us calculate $C_{\textbf{K},0}$. Dashed (solid) lines denote bosons (fermions). The two fermion lines give two Green's functions that need to be summed over the internal momentum and frequency.}
\label{fig:bubble}
\end{figure}

We decouple $H_{int}$ on the surface in the superconducting channel by introducing complex bosonic fields $\Delta_{\textbf{K},i\nu_n}$; see Appendix [\ref{int}] for details. The superconducting instability then corresponds to the divergence of the correlation function $C_{\textbf{K}, i\nu_n} = \langle \overline{\Delta}_{\textbf{K},i\nu_n} \Delta_{\textbf{K}, i\nu_n} \rangle$ at $\textbf{K}=0$, $i\nu_n=0$. Long wavelength equilibrium fluctuations about the mean-field state are subsequently captured by $C_{\textbf{K}, 0}$. In the large $D_S$ limit, $C_{\textbf{K}, i\nu_n}$ is dominated by RPA-like bubble diagrams, which enables a straightforward resummation of the Dyson series, Fig. \ref{fig:Dyson}.
The upshot is
\begin{align}
   C_{\textbf{K}, 0} =  - \frac{U/D_S}{1-\frac{U}{D_S}\chi_{\textbf{K}} }
\end{align}where $-\chi_{\textbf{K}}$ is the bubble shown in Fig. \ref{fig:bubble} 
and given by
\begin{align}
\chi_{\textbf{K}} = \frac{1}{\beta}\sum_{i\omega_n} \intop_{\textbf{k} }\text{Tr} 
\bigg[g^T_{\textbf{k}+\frac{\textbf{K}}{2}, i\omega_n }g_{\textbf{k}-\frac{\textbf{K}}{2}, -i \omega_n}\bigg] \Theta( \omega_D - |i \omega_n|) 
\label{eq:chi-K}
\end{align}
Here, we have introduced a phenomenological Debye frequency $\omega_D$ to model conventional, phonon-mediated pairing.  The superconducting instability now corresponds to the condition $\chi_{0} = D_S/U$.

If $g_{\textbf{k},i\omega_n}$ were the electron Green's function in a conventional metal, its only non-analyticity would have been simple poles on the real axis. For the surface of a Weyl semimetal, the Green's function also has branch cuts on the real axis, so the Matsubara sum must be done with greater care. We carry out this exercise for an explicit model below. Nonetheless, the branch cuts do not change the result qualitatively in meaningful limits.

\section{Tractable layered model} \label{layeredmodel}

We consider a minimal model consisting of alternating layers of spinful electron and hole metals with dispersion $\pm\xi_{\sigma ,\textbf{k}}-\mu$ stacked along $z$ and alternating, real interlayer couplings $t_{\sigma ,\textbf{k}}, -t^\prime_{\sigma ,\textbf{k}}$. Its second-quantized Hamiltonian for an $L$-layered slab is given by
\begin{align}
H = & \intop_{\textbf{k}} \sum_{z=0}^{L-1}\sum_{\sigma=\uparrow,\downarrow}[(-1)^{z} \xi_{\sigma ,\textbf{k}}-\mu] c_{\sigma, \textbf{k} ,z}^\dagger c_{\sigma, \textbf{k} ,z}  +\\  \nonumber
& \left[\cos^2\left(\frac{\pi z}{2}\right)t_{\sigma ,\textbf{k} } - \sin^2\left(\frac{\pi z}{2}\right)t^\prime_{\sigma ,\textbf{k} }\right]c_{\sigma, \textbf{k}, z}^\dagger c_{\sigma, \textbf{k}, z+1} + h.c. \\ \nonumber
\end{align}
where $c_{\sigma, \textbf{k}, z}^\dagger$ creates an electron with spin $\sigma$ at layer $z$ and 2D momentum $\textbf{k} = (k_x, k_y)$. The model clearly conserves spin and has two layers in each unit cell. Its bulk Bloch Hamiltonian in the bilayer basis in the $\sigma$ sector is
\begin{equation}
    H_{\sigma,\textbf{k}} = \begin{pmatrix}
        \xi_{\sigma,\textbf{k}}-\mu & t_{\sigma,\textbf{k}} - t^\prime_{\sigma,\textbf{k}}e^{-2ik_zc} \\
        t_{\sigma,\textbf{k}} - t^\prime_{\sigma,\textbf{k}}e^{2ik_zc} & -\xi_{\sigma,\textbf{k}}-\mu
    \end{pmatrix}
\end{equation}
where $c$ is the interlayer spacing, assumed constant within and between unit cells for simplicity. The interlayer terms are phenomenologically chosen to produce Fermi arcs on the $z=0$ surface along $\xi_{\sigma ,\textbf{k}}=\mu$ when $t^2_{\sigma ,\textbf{k}} < t^{\prime 2}_{\sigma ,\textbf{k}}$. This results in bulk Weyl nodes in the $k_z=0$ plane whenever $t_{\sigma ,\textbf{k}}= t^\prime_{\sigma,\textbf{k}}$. Near the $j^{th}$ Weyl node in the $\sigma$ sector, at $(\textbf{k},k_z) = (\textbf{K}_{\sigma,j},0)$, the low energy Hamiltonian can be written as
\begin{equation}
    H^{\text{Weyl}}_{\sigma,j} = (\textbf{v}_{\sigma,j}\cdot\textbf{p})\tau_z + (\textbf{u}_{\sigma,j}\cdot\textbf{p})\tau_x + (w_{\sigma,j} p_z) \tau_y -\mu
    \label{eq:H-Weyl}
\end{equation}
where $\tau_i$ are Pauli matrices in the bilayer basis, $(\textbf{p},p_z)$ is the 3D momentum relative to the Weyl node and $\textbf{v}_{\sigma,j}=\left.\boldsymbol{\nabla}_{\textbf{k}}\xi_{\sigma,{\textbf{k}}}\right|_{\textbf{k}=\textbf{K}_{\sigma,j}}$, $\textbf{u}_{\sigma,j}=\left.\boldsymbol{\nabla}_\textbf{k}(t_{\sigma,\textbf{k}}-t^\prime_{\sigma,\textbf{k}})\right|_{\textbf{k}=\textbf{K}_{\sigma,j}}$, $w_{\sigma,j} = -2t^\prime_{\sigma,\textbf{K}_{\sigma,j}}c$ are Weyl velocities.

\begin{figure}[h]
\includegraphics[width=0.9\columnwidth]{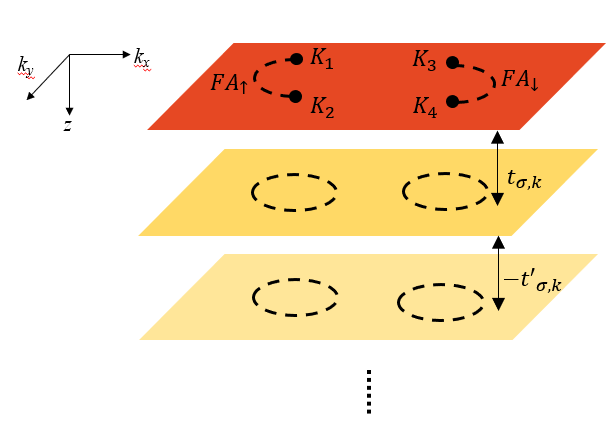}
\caption{\label{fig:model} Minimal layered model of a time-reversal symmetric Weyl semimetal showing two Fermi arcs with opposite spins. }
\end{figure}

\begin{figure}[h]
\includegraphics[width=0.9\columnwidth]{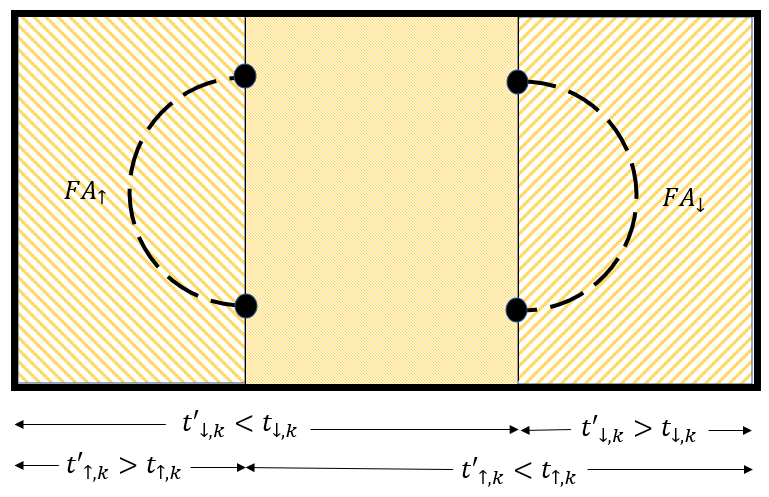}
\caption{\label{fig:surfacelayer} Surface layer of Weyl semimetal with Fermi arcs. The Fermi arcs form when $t'_{\sigma, \textbf{k}} > t_{\sigma, \textbf{k}}$ and $\xi_{\sigma, \textbf{k}} = 0$}.  
\end{figure}

For this model, $g_{\textbf{k},i\omega_n}$ can be calculated analytically in the semi-infinite limit, $L \rightarrow\infty$, following \cite{Hosur2012a}. It is a $2\times2$ diagonal matrix in the spin basis given by
\begin{align}
    g_{\sigma \sigma',\textbf{k}, i\omega_n} & =  \delta_{\sigma\sigma'}\frac{a_{\sigma,\textbf{k}, i\omega_n} + \sqrt{b^+_{\sigma, \textbf{k}, i\omega_n} b^-_{\sigma,\textbf{k}, i\omega_n}} }{2 t'^2_{ \sigma,\textbf{k}}(i\omega_n  + \mu - \xi_{\sigma,\textbf{k}})} \\
    a_{\sigma,\textbf{k}, i\omega_n} &= (i\omega_n +\mu) ^2 - \xi^2_{ \sigma, \textbf{k}} - t^2_{\sigma ,\textbf{k}} + t^{\prime 2}_{\sigma,\textbf{k}} \nonumber \\
    b^{\pm}_{\sigma,\textbf{k}, i\omega_n}  &= (i\omega_n + \mu) ^2 - E^{\pm^2}_{\sigma,\textbf{k}} \nonumber\\ 
    E^{\pm}_{\sigma,\textbf{k}}  &=  \sqrt{\xi^2_{ \sigma,\textbf{k}}  + \left(t_{\sigma,\textbf{k}} \pm t'_{\sigma,\textbf{k}}\right)^2}\\ \nonumber
\end{align}
$g_{\textbf{k},\omega}$ has non-analyticities on the real frequency axis in the form of poles at $\omega =  \mu  +\xi_{\sigma, \textbf{k}}$ that represent the Fermi arcs when $\omega=0$, and a pair of square root branch cuts defined by $E^-_{\sigma, \textbf{k}} < |\omega + \mu|< E^+_{\sigma, \textbf{k}}$ that corresponds to $\omega$ being inside the bulk conduction and valence bands and capture the projection of these bands onto the surface. Along $\xi_{\sigma, \textbf{k}} = \mu$, the surface also carries Luttinger arcs, defined as zeros of $\det(g_{\textbf{k},0})$, that form closed loops with the Fermi arcs when $\mu=0$ \cite{Obakpolor2022}.

This model is a variant of the spinless model introduced in Ref. \cite{Hosur2012a}. Here, we assume two decoupled copies of the model, one for each spin, and ensure time-reversal symmetry by requiring $t_{\sigma,\textbf{k}}$, $t^\prime_{\sigma,\textbf{k}}$ and $\xi_{\sigma,\textbf{k}}$ to be unchanged under the simultaneous reversal of spin and momentum, $\sigma\to-\sigma,\textbf{k}\to-\textbf{k}$. It contains a single orbital degree of freedom in each layer, $D_z=1$ $\forall$ $z$, so we will suppress the index $n_z$ henceforth. We also suppress the spin index below for brevity and assume all functions to be the ones for spin-up, i.e., $\xi_{\textbf{k}}\equiv\xi_{\sigma,\textbf{k}}$, etc.

\section{Surface instability}
 
We now use the above Green's function to evaluate $\chi_{0}$ in Eq. (\ref{eq:chi-K}) to obtain the instability. The trace over spin simply gives a factor of 2. The pair of Green's functions yields two poles, at $\omega=\xi_{\textbf{k}+\textbf{K}/2}-\mu,-\xi_{\textbf{k}-\textbf{K}/2}-\mu$, and four branch cuts, defined by $E^-_{\textbf{k}+\textbf{K}/2} < |\pm\omega + \mu|< E^+_{\textbf{k}+\textbf{K}/2}$. Branch cuts from one Green's function factor can overlap with poles and branch cuts from the other, so the frequency integrals must be performed carefully. Summing over Matsubara frequencies gives separate contributions from the poles and branch cuts of $g_{\textbf{k},\omega}$, $\chi_{0}  = \chi_\text{FA} + \chi_\text{proj}$. 

\subsection{Fermi arc contribution} \label{FermiArc}

The first contribution is
\begin{align} \label{bigeq}
\chi_\text{FA} =& \intop_{\textbf{k}} \tanh\left(\frac{\xi_\textbf{k}-\mu}{2T}\right)R\left(1-\frac{t^2_\textbf{k}}{t^{\prime2}_\textbf{k}}\right)\frac{\Theta(\omega_D - |\xi_\textbf{k} - \mu|)}{\xi_\textbf{k}-\mu} \nonumber\\
\times&\frac{t^{\prime2}_{\textbf{k}}-t^2_{\textbf{k}} + 4\mu(\mu -\xi_{\textbf{k}}) + \sqrt{
\prod \limits_{\lambda=\pm}(2 \mu - \xi_{\textbf{k}})^2 - \left(E_\textbf{k}^{\lambda}\right)^2}}{2t'^2_{\textbf{k}}}
\end{align}
where $R(x)=(x+|x|)/2$ is the ramp function. For $\omega_D\ll$ the hopping energy scales, it is useful to work in momentum coordinates $(k_\parallel,k_\perp)$ parallel and perpendicular to the contour $\xi_\textbf{k}=\mu$. Near this contour, we can approximate $\xi_\textbf{k}=\mu+v_{k_\parallel}k_\perp$. This turns the above expression into a sum of integrals around each Fermi arc, $\chi_\text{FA}=\sum_i \chi_{\text{FA}_i}$, with
\begin{equation}
    \chi_{\text{FA}_i}=\intop_{k_\parallel\in\text{FA}_i}R^2\left(1-\frac{t^2_{k_\parallel}}{t^{\prime2}_{k_\parallel}}\right)\intop_{|k_\perp|<\frac{\omega_D}{|v_{k_\parallel}|}}\frac{\tanh[v_{k_\parallel}k_\perp/2T]}{v_{k_\parallel}k_\perp}
\end{equation}
For $\omega_D\gg T$, the $k_\perp$-integral is dominated by the region $2T<|v_{k_\parallel}k_\perp|<\omega_D$ where $|\tanh[v_{k_\parallel}k_\perp/2T]|\approx1$ and evaluates to $(1/\pi |v_{k_\parallel}|)\ln(\omega_D/2T)$. As a result,
\begin{align}
    \chi_{\text{FA}_i}\approx&\ln\left(\frac{\omega_D}{2T}\right)\intop_{k_\parallel\in\text{FA}_i}\frac{1}{\pi |v_{k_\parallel}|}R^2\left(1-\frac{t^2_{k_\parallel}}{t^{\prime2}_{k_\parallel}}\right)\nonumber\\
    \approx&\frac{l_{\text{FA}_i}}{2\pi^2}\ln\left(\frac{\omega_D}{2T}\right)\left<\frac{1}{|v|}\right>_{\text{FA}_i}
    \label{eq:chi-FAi}
\end{align}
where $l_{\text{FA}_i}$ is the length of the $i^{th}$ Fermi arc and $\left<\dots\right>_{\text{FA}_i}$ denotes a weighted average over this Fermi arc with $k_\parallel$ dependent weight $R^2(1-t^2_{k_\parallel}/t^{\prime2}_{k_\parallel})$. Eq. (\ref{eq:chi-FAi}) matches the corresponding result for a 2D metal if $l_{\textrm{FA}_i}$ is replaced by the perimeter of the Fermi surface and the weight is $k_\parallel$ independent. Thus, Fermi arcs behave like a metallic Fermi surface for harbouring a Cooper instability.

\subsection{Contribution from bulk states}\label{sec:bulk projection}

Next, we evaluate $\chi_\text{proj}$, the contribution to $\chi_0$ from the projection of the bulk states onto the surface, captured by the branch cuts in $g_{\textbf{k},\pm\omega}$. Explicitly, we find
\begin{align}
    \chi_\text{proj} &= -2\intop_{\textbf{k}} \intop_{\omega\in\text{BC}}
    \tanh\left(\frac{\omega}{2T}\right)\sqrt{\left|b^+_{\textbf{k}, \omega}b^-_{\textbf{k}, \omega}\right|}\text{sgn}(\omega+\mu)\nonumber\\
    &\times\frac{a_{-\omega}+\sqrt{R\left[b^+_{\textbf{k},-\omega}b^-_{\textbf{k},-\omega}\right]}}{t'^4_{\textbf{k}} \left[\omega^2 - (\mu -  \xi_{\textbf{k}})^2\right]}\Theta( \omega_D - | \omega|)
    \label{eq:chi-proj-general}
\end{align}
where $\omega\in$ BC denotes the branch cut region $E^-_{\textbf{k}} < |\omega + \mu|< E^+_{\textbf{k}}$ and the factor of $\text{sgn}(\omega+\mu)$ comes from selecting the principal values of the square roots.

In the  regime, $\omega_D  \ll E^+_{\textbf{k}}$, the  conditions $\pm\omega\in\text{BC}$ reduce to $E_{\textbf{k}}^-<|\pm\omega+\mu|$. Physically, this ensures that $\chi_\text{proj}$ receives contributions only from $\mathbf{k}$-space regions defined by surface projections of bulk Fermi surfaces enclosing the Weyl nodes. Hence, we can linearize around the Weyl points as $\xi_{\textbf{p}} \approx \textbf{v}_j \cdot \textbf{p}$, $t_{\textbf{p}}\approx t_j+\textbf{u}_j \cdot \textbf{p}/2$ and $t^\prime_{\textbf{p}} \approx t_j-\textbf{u}_j \cdot \textbf{p}/2$. Then, $E^-_{\textbf{k}}\approx \sqrt{(\textbf{v}_j \cdot \textbf{p})^2+(\textbf{u}_j \cdot \textbf{p})^2}\equiv\epsilon_{\textbf{p}}$, 
$a_{\textbf{k},\omega}\approx -2t_j\textbf{u}_j \cdot \textbf{p}$, $b^-_{\textbf{k},\omega}\approx(\omega+\mu)^2-\epsilon_{\textbf{p}}^2$ and $b^+_{\textbf{k},\omega}\approx-4t_j^2$ near the $j^{th}$ node.  $\chi_\text{proj}$ can then be written as $\chi_\text{proj}=\sum_j \chi_{\text{proj},j}$ where 
\begin{align}
    \chi^\text{proj}_j &= \frac{2}{\pi t_j^2}\intop_{0}^{\omega_D}d\omega\tanh\left(\frac{\omega}{2T}\right)\times\\
    &\intop_{\left|\omega-|\mu|\right|<\epsilon_\textbf{p}<\omega+|\mu|} \frac{\sqrt{\left[(\omega+|\mu|)^2-\epsilon_{\textbf{p}}^2\right]\left[\epsilon_{\textbf{p}}^2-(\omega-|\mu|)^2\right]}}{(|\mu| -  \textbf{v}_j \cdot \textbf{p})^2 - \omega^2}\nonumber
    \label{eq:chi-proj-general}
\end{align}

The pseudo-relativistic form of $\epsilon_{\textbf{p}}$ makes the $\textbf{p}$ integral analytically tractable but rather unwieldy. The complications can be avoided by assuming $\textbf{u}_j\perp\textbf{v}_j$ at the cost of $O(1)$ prefactors. Under this assumption, 
\begin{align}
    \chi^\text{proj}_j &= \frac{8|\mu|^{3/2}}{3\pi^2 t_j^2|u_j v_j|}\intop_{0}^{\omega_D)}d\omega\tanh\left(\frac{\omega}{2T}\right)\sqrt{\omega}\nonumber\\
    & \approx
 \left(\frac{4}{3\pi t_j}\right)^2\frac{ |\mu|^{3/2}\omega_D^{3/2}}{|u_j v_j|}
\end{align}

Importantly, this is a small number compared to $\chi_\text{FA}$ as it is suppressed by powers of $\mu/t_j$ and $\omega_D/t_j$. Thus, $\chi_0\approx\chi_\text{FA}$, and the surface instability is determined mainly by the FAs and resembles that of an ordinary 2D metal. The transition temperature follows from setting $\chi_0=1/U$. Explicitly,
\begin{equation}
    T_C^\text{surf} = \frac{\omega_D}{2}\exp\left[-\frac{2\pi^2}{U\sum_i l_{\text{FA}_i}\left<\frac{1}{|v|}\right>_{\text{FA}_i}}\right]
\end{equation}
Strictly speaking, this is expected to be a Berzinskii-Kosterlitz-Thouless transition rather than a true superconducting transition.

\section{Bulk instability} \label{bulk instability}

We now study the superconducting instability in the bulk. We begin with the Hamiltonian near a Weyl node, Eq. (\ref{eq:H-Weyl}), and compute the appropriate susceptibility $\chi_\text{bulk}$ following the procedure described in Sec. \ref{sec:formalism}. We continue to suppress the spin index to avoid notational clutter, use spin-up functions only (e.g. $\xi_\textbf{k}\equiv\xi_{\sigma,\textbf{k}}$ etc.) and note that the spin sum merely contributes a factor of 2 to $\chi_\text{bulk}$.

The Green's function near the $j^{th}$ Weyl node is 
\begin{align}
    G_{j,\textbf{p},i\omega_n} = \frac{1}{i\omega_n+\mu -\textbf{v}_{j}\cdot\textbf{p}\tau_z  - \textbf{u}_{j}\cdot\textbf{p}\tau_x - w_{j} p_z \tau_y}
\end{align}
Thus, $\chi_{\text{bulk},j}=2T\sum_{i\omega_n}\intop_{\textbf{p}}\text{Tr}\left[G_{j,\textbf{p},i\omega_n}^TG_{j,\textbf{p},-i\omega_n}\right]\Theta(\omega_D-|i\omega_n|)$ is given by
\begin{align}
    \chi^{\text{bulk}}_j&=2T\sum_{i\omega_n}\Theta(\omega_D-|i\omega_n|) \times\nonumber\\
    &\intop_{\textbf{p}}\frac{\mu^2-(i\omega_n)^2 + \varepsilon_{\textbf{p},p_z}^2 - 2(w_{j} p_z)^2}{\prod_{s=\pm}\left[(i\omega_n+s\mu)^2-\varepsilon_{\textbf{p},p_z}^2\right]}
    \label{eq:chi-bulk-def}
\end{align}
where $\varepsilon_{\textbf{p},p_z} = \sqrt{\epsilon_{\textbf{p}}^2+(w_jp_z)^2}$ is the pseudo-relativistic dispersion near the Weyl node. Once again, the integrals are analytically tractable in this limit and yield $\chi^\text{bulk}=\sum_j\chi^\text{bulk}_j$ with
\begin{equation}
    \chi^\text{bulk}_j = \frac{\mu^2}{3\pi^2|(\textbf{u}_j\times\textbf{v}_j)\cdot \textbf{w}_j|}\ln\left(\frac{\omega_D}{2T}\right)
\end{equation}
for $\omega_D\gg T$.

The bulk transition temperature is given by
\begin{equation}
    T_C^\text{bulk} = \frac{\omega_D}{2} \exp\left[- \frac{3\pi^2}{U \mu^2\sum_j\frac{1}{|(\textbf{u}_j\times\textbf{v}_j)\cdot \textbf{w}_j|}}\right]
\end{equation}
Naturally, $T_C^\text{bulk}\to0$ as $\mu=0$ and grows smoothly with $\mu$. Crucially, there exists a parametrically large regime in which $T_C^\text{bulk} < T_C^\text{surf}$, namely,
\begin{equation}
    {\sum_j\frac{\mu^2}{3|(\textbf{u}_j\times\textbf{v}_j)\cdot \textbf{w}_j|}} < {\sum_i l_{\text{FA}_i}\left<\frac{1}{2|v|}\right>_{\text{FA}_i}}
\end{equation}For temperatures between these two values, the surface will superconduct, and the bulk will remain metallic with the caveat that the surface superconductivity will presumably be of Berezinskii-Kosterlitz-Thouless type.

\section{Experimental relevance} \label{experiment}

This result is pertinent to the observations of superconductivity in t-PtBi\textsubscript{2}. In particular, Schimmel et al. saw a wide range of surface superconducting gaps in the tunneling spectrum of t-PtBi\textsubscript{2}, with the largest gaps corresponding to $T_C\sim 100$K range \cite{Schimmel2023}. In comparison, transport measurements in bulk crystals displayed $T_C\sim0.6$K \cite{Shipunov2020}. The authors of Ref. \cite{Schimmel2023} speculated that the higher $T_C$ in tunneling measurements was due to a transition from bulk to surface superconductivity and asked whether the surface superconductivity is connected to the topologically non-trivial states found on the surface of a Type-I Weyl semimetal. We have shown, in a toy model, that the dominant part of surface superconductivity indeed emerges from FA states and yields a higher $T_C$ on the surface than in the bulk.

Our work  is also relevant to the observation of superconductivity with $T_C$ of 6K - 9K in powdered samples of NbP, a Type-I Weyl semimetal, by Baenitz et al. \cite{Baenitz2019}. This was in agreement with another published value, $T_C\sim$7.5K \cite{Kumar2016}, for powdered NbP compounds. Baenitz et al. reported a superconducting fraction of only 6.2 \% and gave two possible explanations based on grain size effects for the small fraction. The first one involved strain on the grains, turning the material into a Type-II Weyl semimetal, which has a bigger Fermi surface and is thus more likely to superconduct. The second explanation involved superconductivity developing on the surface, which can lead to a sizeable signature in powder samples. We have shown that the second latter picture is possible, at least within mean-field theory in a phenomenological model. Moreover, we note that in the first picture, different samples would likely turn into Type-II Weyl semimetals with differing sizes of electron and hole pockets and exhibit vastly different $T_C$, unlike what was observed. In contrast, intrinsic surface superconductivity is more likely to yield similar transition temperatures in different samples. Our picture can be tested by studying superconductivity in bulk and thin films of NbP. If superconductivity intrinsically occurred on the surface, thin films would display a larger superconducting fraction than bulk crystals, in sharp contrast to the behavior of conventional metallic superconductors.

\section{Summary}

We have shown that under a mean-field limit in a phenomenological model of a TWSM, there exists a parametrically large regime where the surface has a superconducting instability, whereas the bulk remains in the normal state. Furthermore, we find that the instability is governed mainly by the Fermi arc surface states, and the contribution from the surface projection of the bulk Fermi surface is negligible. This result pertains to recent experiments on NbP and t-PtBi\textsubscript{2} that raised the possibility of intrinsic surface superconductivity in TWSMs.

\begin{acknowledgments}
We acknowledge financial support from the National Science Foundation grant no. DMR 2047193. We are grateful to Kai Chen and Osakpolor Obakpolor for useful discussions.
\end{acknowledgments}

\begin{widetext}

\appendix
\section{Surface Green's Function}
\label{sgf}
Let $H_{\mathbf{k}}^{B}$ denote the Bloch Hamiltonian of an $L$-layered
time-reversal symmetric system that has $2D_{z}$ degrees of freedom
in the $z^{th}$ layer. Time-reversal symmetry (TRS) ensures that
each layer has an even number of degrees of freedom. The layers are
unrelated in general but repeat periodically in lattice models. Now,
let us add a layer at $z=0$ that we refer to as the ``surface''.
The Hamiltonian for the full system is of the form
\begin{equation}
H_{\mathbf{k}}=\left(\begin{array}{cc}
H_{\mathbf{k}}^{S} & h_{\mathbf{k}}\\
h_{\mathbf{k}}^{\dagger} & H_{\mathbf{k}}^{B}
\end{array}\right)
\end{equation}
We will use $\bar{b},b$ to denote Grassman variables for fermions
in layers $1\dots L$, and $\bar{s},s$ for surface fermions. Contractions
over layers ($z$), orbitals ($n_{z}$), and spin ($\sigma$) will
be denoted by ``$\cdot$'' while integrals will be written in shorthand
as $\intop_{\mathbf{k},\tau}\equiv\intop_{0}^{\beta}d\tau\int\frac{d^{2}k}{(2\pi)^{2}}$.
In this notation, the Euclidean path integrals for the $L$- and $(L+1)$-layered
systems are $Z_{0}^{B}=\int\mathcal{D}\left[\bar{b},b\right]\exp\left[-\mathcal{S}_{0}^{B}\left(\bar{b},b\right)\right]$
and $Z_{0}=\int\mathcal{D}\left[\bar{b},b,\bar{s},s\right]\exp\left[-\mathcal{S}_{0}\left(\bar{b},b,\bar{s},s\right)\right]$
where
\begin{align}
\mathcal{S}_{0}\left(\bar{b},b\right) & =-\intop_{\mathbf{k},\tau}\sum_{z,z'=1}^{L}\sum_{\sigma,\sigma'\in\uparrow,\downarrow}\bar{b}_{\mathbf{k},z,n_{z},\sigma}\left(\partial_{\tau}\delta_{(z,n_{z},\sigma),(z',n_{z'},\sigma')}+H_{\mathbf{k},(z,n_{z},\sigma),(z',n_{z'},\sigma')}^{B}\right)b_{\mathbf{k},z',n_{z'},\sigma'}\\
 & \equiv\intop_{\mathbf{k},\tau}\bar{b}_{\mathbf{k}}\cdot\left[G_{\mathbf{k}}^{B}(\tau)\right]^{-1}\cdot b_{\mathbf{k}}\\
\mathcal{S}_{0}\left(\bar{b},b,\bar{s},s\right) & =-\intop_{\mathbf{k},\tau}\left(\bar{s}_{\mathbf{k}},\bar{b}_{\mathbf{k}}\right)\cdot\left(\partial_{\tau}+H_{\mathbf{k}}\right)\cdot\left(\begin{array}{c}
s_{\mathbf{k}}\\
b_{\mathbf{k}}
\end{array}\right)\\
 & \equiv\intop_{\mathbf{k},\tau}\left(\bar{s}_{\mathbf{k}},\bar{b}_{\mathbf{k}}\right)\cdot\left[G_{\mathbf{k}}(\tau)\right]^{-1}\cdot\left(\begin{array}{c}
s_{\mathbf{k}}\\
b_{\mathbf{k}}
\end{array}\right)
\end{align}
We have introduced imaginary time Green's functions $G_{\mathbf{k}}^{B}(\tau)$
and $G_{\mathbf{k}}(\tau)$ for the $L$- and $(L+1)$-layered system.
Integrating out the $b$-fermions yields an effective surface Green's
function $g_{\mathbf{k}}(\tau)$ as follows:
\begin{align}
Z_{0}^{S} & =\frac{Z_{0}}{Z_{0}^{B}}\equiv\int\mathcal{D}\left[\bar{s},s\right]\exp\left[-\mathcal{S}_{0}^{S}\left(\bar{s},s\right)\right]\\
\mathcal{S}_{0}^{S}\left(\bar{s},s\right) & =-\intop_{\mathbf{k},\tau}\bar{s}_{\mathbf{k}}\cdot\left[\partial_{\tau}+H_{\mathbf{k}}^{S}+h_{\mathbf{k}}G_{\mathbf{k}}^{B}(\tau)h_{\mathbf{k}}^{\dagger}\right]\cdot s_{\mathbf{k}}\\
\implies g_{\mathbf{k}}(\tau) & =-\left(\partial_{\tau}+H_{\mathbf{k}}^{S}+h_{\mathbf{k}}G_{\mathbf{k}}^{B}(\tau)h_{\mathbf{k}}^{\dagger}\right)^{-1}
\end{align}
The Matsubara Green's functions $G_{\mathbf{k}}^{B}(i\omega_{n})$,
$G_{\mathbf{k}}(i\omega_{n})$ and $g_{\mathbf{k}}(i\omega_{n})$
can be obtained straightforwardly by the replacement $\partial_{\tau}\to-i\omega_{n}$
in the above equations.

\section{Interaction}
\label{int}
Since the interaction is local, the path integral for the
full interacting system factorizes between the bulk and the surface:
$Z=Z^{B}Z^{S}$ where 
\begin{align}
Z^{B} & =\int\mathcal{D}\left[\bar{b},b\right]\exp\left[-\mathcal{S}_{0}^{B}\left(\bar{b},b\right)-\mathcal{S}_{int}^{B}\left(\bar{b},b\right)\right]\\
\mathcal{S}_{int}^{B}\left(\bar{b},b\right) & =-\intop_{\mathbf{K},\tau}\sum_{z}\frac{U}{D_{z}}\bar{\mathbb{B}}_{\mathbf{K},z}\mathbb{B}_{\mathbf{K},z}\\
\mathbb{B}_{\mathbf{K},z} & =\sum_{n_{z}=1}^{D_{z}}\intop_{\mathbf{k}}b_{\frac{\mathbf{K}}{2}+\mathbf{k},z,n_{z}\downarrow}b_{\frac{\mathbf{K}}{2}-\mathbf{k},z,n_{z},\uparrow}
\end{align}
and 
\begin{align}
Z^{S} & =\int\mathcal{D}\left[\bar{s},s\right]\exp\left[-\mathcal{S}_{0}^{S}\left(\bar{s},s\right)-\mathcal{S}_{int}^{B}\left(\bar{s},s\right)\right]\\
\mathcal{S}_{int}^{S}\left(\bar{s},s\right) & =-\frac{U}{D_{S}}\intop_{\mathbf{K},\tau}\bar{\mathbb{S}}_{\mathbf{K}}\mathbb{S}_{\mathbf{K}}\\
\mathbb{S}_{\mathbf{K}} & =\sum_{n_{S}=1}^{D_{S}}\intop_{\mathbf{k}}s_{\frac{\mathbf{K}}{2}+\mathbf{k},n_{S}\downarrow}s_{\frac{\mathbf{K}}{2}-\mathbf{k},n_{S},\uparrow}
\end{align}The fermion bilinears $\mathbb{B}_{\mathbf{K},z}$ and $\mathbb{S}_{\mathbf{K}}$
are bosonic variables, and $D_{S}\equiv D_{0}$ is the number of degrees
of freedom in the $z=0$ surface layer.

To investigate surface superconductivity, we focus on $Z^{S}$. Decoupling
the interaction term in the $s$-wave pairing channel through another
bosonic field $\Delta_{2\mathbf{K}}$ gives
\begin{equation}
Z^{S}=\int\mathcal{D}\left[\bar{s},s\right]\exp\left[-\mathcal{S}_{0}^{S}\left(\bar{s},s\right)\right]\int\mathcal{D}\left[\bar{\Delta},\Delta\right]\exp\left[-\mathcal{S}'\left(\bar{\Delta},\Delta,\bar{s},s\right)\right]
\end{equation}
where
\begin{equation}
\mathcal{S}'\left(\bar{\Delta},\Delta,\bar{s},s\right)=\intop_{\mathbf{K},\tau}-\frac{D_{S}}{U}\bar{\Delta}_{\mathbf{K}}\Delta_{\mathbf{K}}+\bar{\mathbb{S}}_{\mathbf{K}}\Delta_{\mathbf{K}}+\bar{\Delta}_{\mathbf{K}}\mathbb{S}_{\mathbf{K}}
\end{equation}

\section{Green's function trace}
In this section, we show how the expressions for $\chi_0$ can be simplified and written as the trace of a product of Green's functions.

\begin{align}
\chi_{0} & =\intop_{\mathbf{k},i\omega_{n}}\left[g_{\mathbf{k}}(i\omega_{n})\right]_{n n^{\prime}}^{\sigma\sigma'}\left[g_{-\mathbf{k}}(-i\omega_{n})\right]_{n n^{\prime}}^{\bar{\sigma}\bar{\sigma}'}\\
 & =\intop_{\mathbf{k},i\omega_{n}}\left[g_{\mathbf{k}}(i\omega_{n})\right]_{n n^{\prime}}^{\sigma\sigma'}\left[\mathcal{T}g_{\mathbf{k}}(i\omega_{n})\mathcal{T}^{-1}\right]_{n n^{\prime}}^{\sigma\sigma'}\\
 & =\intop_{\mathbf{k},i\omega_{n}}\text{tr}\left[g_{\mathbf{k}}^{T}(i\omega_{n})\mathcal{T}g_{\mathbf{k}}(i\omega_{n})\mathcal{T}^{-1}\right]
\end{align}
where the trace runs over both spin and orbital indices, $\intop_{\mathbf{k},i\omega_{n}}=T\sum_{i\omega_{n}}\int\frac{d^{2}k}{(2\pi)^{2}}$
and $\mathcal{T}$ denotes time reversal. We have used the action
of $\mathcal{T}$ on the matrix elements of $g_{\mathbf{k}}(i\omega_{n})$:
\begin{equation}
\left[\mathcal{T}g_{\mathbf{k}}(i\omega_{n})\mathcal{T}^{-1}\right]_{n n^{\prime}}^{\sigma\sigma'}=\left[g_{-\mathbf{k}}(-i\omega_{n})\right]_{n n^{\prime}}^{\bar{\sigma}\bar{\sigma}'}
\end{equation}
and used the identity $\text{tr}\left(AB^{T}\right)=\text{tr}\left(A^{T}B\right)$
to reduce notational clutter. Since the system is $\mathcal{T}$-symmetric,
$\mathcal{T}g_{\mathbf{k}}(i\omega_{n})\mathcal{T}^{-1}=g_{\mathbf{k}}(-i\omega_{n})$.
This gives
\begin{equation}
\chi_{0}=\intop_{\mathbf{k},i\omega_{n}}\text{tr}\left[g_{\mathbf{k}}^{T}(i\omega_{n})g_{\mathbf{k}}(-i\omega_{n})\right]
\end{equation}
Above, we separated the $\sigma$ and $n$ indices for clarity and assumed the orbitals to be $\mathcal{T}$-symmetric. However, the expression in terms of $\text{tr}\left(gg^{T}\right)$ should work even if the orbitals are not $\mathcal{T}$-symmetric. In general:
\begin{align}
\chi_{\mathbf{K}} & =\intop_{\mathbf{k},i\omega_{n}}\left[g_{\mathbf{K}/2+\mathbf{k}}(i\omega_{n})\right]_{n_{S}n_{S}^{\prime}}^{\sigma\sigma'}\left[g_{\mathbf{K}/2-\mathbf{k}}(-i\omega_{n})\right]_{n_{S}n_{S}^{\prime}}^{\bar{\sigma}\bar{\sigma}'}\\
 & =\intop_{\mathbf{k},i\omega_{n}}\left[g_{\mathbf{K}/2+\mathbf{k}}(i\omega_{n})\right]_{n_{S}n_{S}^{\prime}}^{\sigma\sigma'}\left[\mathcal{T}g_{-\mathbf{K}/2+\mathbf{k}}(i\omega_{n})\mathcal{T}^{-1}\right]_{n_{S}n_{S}^{\prime}}^{\sigma\sigma'}\\
 & =\intop_{\mathbf{k},i\omega_{n}}\text{tr}\left[g_{\mathbf{k}+\mathbf{K}/2}^{T}(i\omega_{n})g_{\mathbf{k}-\mathbf{K}/2}(-i\omega_{n})\right]
\end{align}

\section{Integrals for calculating $\chi_{bulk}$}

In this section, we describe the integration steps for computing $\chi_\text{bulk}$. We begin with Eq. (\ref{eq:chi-bulk-def}) from the main text 
\begin{align}
    \chi^\text{bulk}_j=2T\sum_{i\omega_n}\intop_{\textbf{p}}\frac{\mu^2-(i\omega_n)^2 + \varepsilon_{\textbf{p},p_z}^2 - 2(w_{j} p_z)^2}{\prod_{s=\pm}\left[(i\omega_n+s\mu)^2-\varepsilon_{\textbf{p},p_z}^2\right]\Theta(\omega_D-|i\omega_n|)}
\end{align}
where $\varepsilon_{\textbf{p},p_z} = \sqrt{\epsilon_{\textbf{p}}^2+(w_jp_z)^2}$ and $\epsilon_{\textbf{p}}=\sqrt{(\textbf{v}_j \cdot \textbf{p})^2+(\textbf{u}_j \cdot \textbf{p})^2}$ are massless relativistic dispersions in 3D and 2D. To bring the integrals into a spherically symmetric form, we rotate and rescale the momenta as
\begin{equation}
\left(\begin{array}{c}
q_x\\
q_y\\
q_z
\end{array}\right)=\left(\begin{array}{ccc}
V_j & 0 & 0\\
0 & U_j & 0\\
0 & 0 & w_j
\end{array}\right)\left(\begin{array}{ccc}
\cos\theta_j & -\sin\theta_j & 0\\
\sin\theta_j & \cos\theta_j & 0\\
0 & 0 & 1
\end{array}\right)\left(\begin{array}{c}
p_{\parallel}\\
p_{\perp}\\
p_z
\end{array}\right)\label{eq:transformations-3D}
\end{equation}
This gives,
\begin{align}
    \chi^\text{bulk}_j&=2T\sum_{i\omega_n}\int\frac{d^3q}{U_jV_jw_j(2\pi)^3}\frac{\mu^2-(i\omega_n)^2 + q_x^2+q_y^2-q_z^2}{\prod_{s=\pm}\left[(i\omega_n+s\mu)^2-q^2\right]}\Theta(\omega_D-|i\omega_n|)\nonumber\\
    &=\frac{1}{\pi^2U_jV_jw_j}T\sum_{i\omega_n}\intop_0^\infty q^2dq\frac{\mu^2-(i\omega_n)^2 + q^2/3}{\left[(i\omega_n+\mu)^2-q^2\right]\left[(i\omega_n-\mu)^2-q^2\right]}\Theta(\omega_D-|i\omega_n|)
\end{align}
Performing the Matsubara sum and some algebra gives
\begin{align}
    \chi^\text{bulk}_j&=-\frac{1}{4\pi^2U_jV_jw_j\mu}\intop_{\mu-\omega_D}^{\mu+\omega_D} q^2dq \frac{q/3-\mu}{q-\mu}\tanh\left(\frac{q-\mu}{2T}\right)
\end{align}
Shifting $q$ by $\mu$ results in a symmetric integration range and causes several terms to vanish. We are then left with
\begin{align}
    \chi^\text{bulk}_j&=\frac{\mu^2}{6\pi^2U_jV_jw_j}\intop_{-\omega_D}^{\omega_D} dq \frac{\tanh(q/2T)}{q}
\end{align}
For $\omega_D\gg T$,
\begin{align}
    \chi^\text{bulk}_j\approx\frac{\mu^2}{3\pi^2|(\textbf{u}_j\times\textbf{v}_j)\cdot \textbf{w}_j|}\left[\ln\left(\frac{\omega_D}{2T}\right)+O(1)\right]
\end{align}

\end{widetext}

\bibliography{library}

\end{document}